\documentclass[a4paper,11pt]{article}
\usepackage{amsmath}
\usepackage{amsfonts}
\usepackage{graphicx}
\usepackage{amssymb}
\usepackage{epsfig}
\usepackage{empheq}
\usepackage{filecontents}

\usepackage{hyperref}

\def\ba{\begin{eqnarray}}
\def\ea{\end{eqnarray}}
\def\be{\begin{equation}}
\def\ee{\end{equation}}

\def\nn{\nonumber}

\def\O{\mathcal{O}}

\def\L{\mathcal{L}}

\def\Z{\mathcal{Z}}
\def\G{\mathcal{G}}

\def\d{\partial}

\def\M{\mathcal{M}}

\usepackage{tikz}
\usetikzlibrary{arrows,shapes}
\usetikzlibrary{trees}
\usetikzlibrary{matrix,arrows} 				% For commutative diagram
\usetikzlibrary{positioning}				% For "above of=" commands
\usetikzlibrary{calc,through}				% For coordinates
\usetikzlibrary{decorations.pathreplacing}  % For curly braces
\usepackage{pgffor}							% For repeating patterns

\usetikzlibrary{decorations.pathmorphing}	% For Feynman Diagrams
\usetikzlibrary{decorations.markings}

\numberwithin{equation}{section}

\begin{document}

\begin{titlepage}

\vskip 1.5cm

Nikhef-2021-029

\vskip 1.5cm

\begin{center}
{\LARGE \bf{Finite Callan-Symanzik renormalisation for multiple scalar fields}}
\vskip 1.5cm

\large{{\bf Sander Mooij$^{1,2}$ and Mikhail Shaposhnikov$^2$}}
\vskip 1cm

\emph{Nikhef, Theory Group \\Science Park 105, \\NL-1098 XG Amsterdam, The Netherlands}

\vskip 0.5cm

\emph{
Institute of Physics, Laboratory for Particle Physics and Cosmology,\\
	\'Ecole Polytechnique F\'ed\'erale de Lausanne, \\CH-1015 Lausanne, Switzerland
	}
	
\vskip 1cm

\tt{sander.mooij@epfl.ch,~mikhail.shaposhnikov@epfl.ch}

\end{center}

\vskip1cm

\begin{center}
\large{{\bf Abstract}}
\end{center}
\vskip 0.5cm

We study a finite, divergence-free approach to renormalisation originally proposed in the early '70s by Blaer and Young, and Callan. It is based on equations similar to the Callan-Symanzik equations and introduced in the context of the $\lambda \phi^4$ theory.  We generalise this method to the case of two interacting scalar fields, with obvious generalisation to an arbitrary number of fields. 

\end{titlepage}

\newpage

\section{Introduction}

There exist formalisms in Quantum Field Theory (QFT) to obtain finite, ``physical" $n$-point correlation functions that do not encounter intermediate ultraviolet (UV) infinities. The most famous example is the R-operation of Bogolubov-Parasuk-Hepp-Zimmermann (see \cite{Bogoliubov} and references therein), which replaces the integrand of the Feynman graphs with an appropriate expression. No UV infinities show up when this expression is integrated. The less-known procedure is based on the Callan-Symanzik (CS) equations \cite{CSCal,CSSym}, usually used for discussion of UV asymptotics of the Green's functions. In \cite{Callan}, on the example of the simplest scalar $\lambda \phi^4$ theory, it was shown that the same equations can be used for the computation of the physical amplitudes without any intermediate infinities. This formalism was extended to quantum electrodynamics in \cite{Blaer}. More recently 't Hooft \cite{tHooft:2004bkn} found a system of equations, different from those of CS, for strongly connected Green's functions.  The solution of this system is free from divergences at any step of computations. 't Hooft noted that the divergence-free methods may allow for a better conceptual understanding of ultraviolet infinities.

Besides infinities, the interesting QFTs may contain widely separated mass scales, such as the electroweak scale and the Grand Unified scale. The Appelquist-Carazzone theorem \cite{Appelquist:1974tg} says that heavy fields decouple at low momenta except for their contribution to renormalization effects. It is interesting to see how this result appears in the finite QFT formalism(s) in which no infinities are present and no renormalisation is needed. 

This paper aims to address this question in the framework of CS formalism\footnote{In  \cite{Mooij:2021ojy}, we employed this method in a discussion of naturalness and the hierarchy problem. }. To do so, one needs a theory with several fields and different masses. The generalisation of the Callan-Symanzik equations to the case of multiple fields has been discussed widely in the literature (see for example \cite{Sirlin,Nish1, Nish2,Kraus}). However, it seems that the ``Callan-Symanzik-inspired" equations governing the CS method for finite renormalisation have only been formulated for the one-(scalar) field case. While a generalisation to fermion fields seems straightforward, handling several fields turns out to be more intricate. In this work, we generalise the CS method to the case of two interacting scalar fields, which, to the best of our knowledge, has never been done.

This paper is organised as follows. In the second Section, we review the original CS method. (A longer introduction can be found in the original lecture notes \cite{Callan}.) Our focus is on the derivation of the equations on which the method relies.  In the third Section, we generalise the argument to the case of two interacting scalar fields. (A further generalisation to $n$ fields is trivial). The fourth Section is devoted to an explicit computation of the one-loop quantum corrections in this two-field model. 

\section{CS method: one scalar field }

\subsection{Formulation: differential equations and boundary conditions}

The CS recipe to (recursively) compute renormalised $n$-point functions $\bar{\Gamma}^{(n)}$ in $\lambda \phi^4$ theory in a manifestly finite way relies on  two differential equations
\ba
2i m^2 ~(1+\gamma)\cdot\bar{\Gamma}^{(n)}_\theta&=& \left[  \left(2m^2 ~\frac{\d}{\d m^2} +\beta~\frac{\d}{\d \lambda}     \right)+n\cdot \gamma\right] \bar{\Gamma}^{(n)}\nn~,\\
2im^2~(1+\gamma)\cdot  \bar{\Gamma}^{(n)}_{\theta \theta} &=& \left(2m^2~\frac{\d}{\d m^2} +\beta~\frac{\d}{\d \lambda}+n\cdot\gamma+\gamma_\theta\right)\bar{\Gamma}^{(n)}_\theta~,
\label{calrel}
\ea
with $\beta$, $\gamma$ and $\gamma_\theta$ for now arbitrary quantities, to be found in the course of solving these equations. $\bar{\Gamma}^{(n)}$ denotes a renormalised, finite $n$-point function\footnote{That is to say, the point of this method is precisely that once the equations above have been established, there is no need to ``renormalise" any $n$-point function. However, to compare with the usual method, and to facilitate a clear derivation of the CS method in the next section, we will still refer to the $\bar{\Gamma}^{(n)}$ as \emph{renormalised} $n$-point functions.} while the quantities $\bar{\Gamma}^{(n)}_\theta$ and $\bar{\Gamma}^{(n)}_{\theta\theta}$ are still to be defined. These equations get supplemented with a set of boundary conditions, which can be divided in tree level conditions
\be
\left[\bar{\Gamma}^{(2)}\right]_{\lambda^0}
= i\left(k^2+m^2\right)~,\qquad
\left[\bar{\Gamma}^{(2)}_\theta\right]_{\lambda^0} = 1~,\qquad
\left[\bar{\Gamma}^{(4)}\right]_{\lambda^1}=-i\lambda~, \label{bountree}
\ee
and boundary conditions
\ba
\left[\frac{d}{dk^2}~\bar{\Gamma}^{(2)}(k^2)\right]_{k^2 = 0} &=&i~, \qquad   \bar{\Gamma}^{(2)}\left(k^2 = 0\right) =i m^2~,
\qquad 
%\nn\\
%\bar{\Gamma}^{(2)}_\theta(k^2=0) &=& 1~,\qquad %\qquad
\bar{\Gamma}^{(4)}\left(k^2 = 0\right) =-i\lambda~.
\label{bounren}
\ea
(Here the notation $\left[\bar{\Gamma}^{(4)}\right]_\lambda$, for example, denotes the order $\lambda$ contribution to the renormalised four-point function $\bar{\Gamma}^{(4)}$.)
From the form of the boundary conditions, it is clear that in this ``finite"\footnote{In this paper, the term ``finite" refers to the absence of \emph{UV} divergences. It does not address IR divergences. Moreover, the CS method needs to be refined in the infrared, because as it stands, it cannot handle massless fields. However, for our ``UV purposes", the current formulation of the theory suffices.} formulation, the CS method leads to the same results for renormalised Green's functions as the traditional zero external momentum subtraction scheme.

The main aim of this paper is to generalise this CS method to the case of two scalar fields. However, to do so, we still need to fill a hole in the one-field analysis. 
Therefore, the goal of this Section is to illustrate their derivation (see also the original papers \cite{Blaer,Callan}). That will serve as a beginning point for the two-field analysis of the next Section.

\subsection{Derivation}

\subsubsection{$\theta$-operation}

Even if the eqs.~(\ref{calrel}) only include renormalised quantities, their derivation begins at the bare level. We begin from the bare Lagrangian
\be
\L_0 = - \frac{1}{2} \d_\mu \phi_0 \d^\mu \phi_0 -\frac{m_0^2}{2}\phi_0^2 - \frac{\lambda_0}{4!}\phi_0^4~.
\ee
The CS approach is all based on the observation that differentiating the (bare) scalar field propagator with respect to $m_0^2$ yields (minus $i$ times) two propagators:
\be
\frac{d}{dm_0^2}\left[ \frac{-i}{k^2+m_0^2}\right] =- i\cdot \left(\frac{-i}{k^2+m_0^2}\right)^2~.
\ee
Obviously, adding an extra propagator to the diagram reduces its degree of divergence by two. Therefore, any two-point diagram can be made finite by two differentiations with respect to the mass parameter. For four-point diagrams, one differentiation is enough. Six-point (and higher) diagrams are convergent by themselves (up to subdivergences, see below). 

Taking this derivative (and multiplying by $-i$) is now denoted as acting with a ``$\theta$-operator" on a propagator. Graphically, it can be depicted as inserting a new ``$\theta$-vertex" on a propagator. This vertex comes with Feynman rule $(-1)$ and effectively cuts the propagator in two. 
Here we will stick to the algebraic representation of the $\theta$-operation:
\be
\Gamma_\theta^{(n)}(k^2) \equiv-i\times \frac{d}{d m_0^2}~ \Gamma^{(n)}(k^2)~. \label{caldef}
\ee
$\Gamma_\theta^{(n)}$ is still divergent for $n=2$. That is why we still need a boundary condition for it, to be formulated in eq.~(\ref{gtcon}).

Finally, bare $\Gamma^{(n)}$ is connected through renormalised $\bar{\Gamma}^{(n)}$ through the usual relation
\be
\Gamma^{(n)} (\lambda_0,m_0) = Z^{n/2}~\bar{\Gamma}^{(n)}(\lambda,m)~. \label{relbar}
\ee
To relate bare $\Gamma^{(n)}_\theta$ to renormalised $\bar{\Gamma}^{(n)}_\theta$ we need to introduce one more object $Z_\theta$:
\be
\Gamma^{(n)}_\theta (\lambda_0,m_0) = Z^{n/2}~Z_\theta ~\bar{\Gamma}_\theta^{(n)}(\lambda,m)~. \label{relbar2}
\ee

\subsubsection{First CS equation}

Now we can just begin from the definition in eq.~(\ref{caldef}) and use eqs.~(\ref{relbar}) and (\ref{relbar2}) to rewrite both sides in terms of renormalised quantities. Decomposing the ``bare" total derivative in terms of ``physical" partial derivatives
\be
\frac{d}{d m_0^2} = \frac{\d m^2}{\d m_0^2} \frac{\d}{\d m^2} + \frac{\d \lambda}{\d m_0^2} \frac{\d}{\d \lambda}~, \label{rewr}
\ee
one obtains
\be
i\cdot\bar{\Gamma}^{(n)}_\theta = Z^{-n/2} Z_\theta^{-1}~\left[ Z^{n/2} \left(  \frac{\d m^2}{\d m_0^2}\frac{\d}{\d m^2} + \frac{\d \lambda}{\d m_0^2}\frac{\d}{\d \lambda}     \right) +\frac{n}{2}~Z^{n/2-1}~\frac{\d Z}{\d m_0^2}\right]\bar{\Gamma}^{(n)} ~,\label{defi}
\ee
and then
\ba
2 im^2  ~\left[ \frac{\d m^2}{\d m_0^2}\right]^{-1}~Z_\theta~\bar{\Gamma}^{(n)}_\theta&=& \Biggl[  \left(2m^2 ~\frac{\d}{\d m^2} +2 m^2 \left[ \frac{\d m^2}{\d m_0^2}\right]^{-1} \frac{\d \lambda}{\d m_0^2}  \frac{\d}{\d \lambda}  \right)\nn\\
&& \qquad \qquad\qquad+n\cdot m^2 \left[ \frac{\d m^2}{\d m_0^2}\right]^{-1}~\frac{\d \ln{Z}}{\d m_0^2}\Biggr] \bar{\Gamma}^{(n)}~. \label{defin}
\ea
We can now eliminate all reference to bare quantities by simply absorbing them into three newly defined, dimensionless quantities:
\ba
G &\equiv & \left[ \frac{\d m^2}{\d m_0^2}\right]^{-1}~Z_\theta~,\nn\\
\beta &\equiv & 2 m^2 \left[ \frac{\d m^2}{\d m_0^2}\right]^{-1} ~\frac{\d \lambda}{\d m_0^2}~,\nn\\
\gamma &\equiv & m^2 \left[ \frac{\d m^2}{\d m_0^2}\right]^{-1}~\frac{\d \ln{Z}}{\d m_0^2}~.  \label{3def}
\ea
It is not difficult to prove \cite{Callan} that $\beta$ must begin at order $\lambda^2$, while $\gamma$ must begin at order $\lambda$. Moreover, they are manifestly convergent in that order. (This can be double-checked from the relations between $m_0^2$ and $m^2$ and between $\lambda_0$ and $\lambda$ in the standard approach.) Finally, $G$ can be found by imposing one more boundary condition
\be
\bar{\Gamma}^{(2)}_\theta(k^2=0)~,\label{gtcon}
\ee
and evaluating eq.~(\ref{defin}) for $n=2$ at $k^2=0$. That gives
\be
G= 1+\gamma~, \label{remov}
\ee
up to all orders. 

With this, we end up with the first equation in eq.~(\ref{calrel}). It contains only finite quantities. Even if we used the notion of bare quantities to derive this equation, we could even just forget about that and take the equation as ``God-given".
Here we have outlined its derivation just to be in a good position to generalise the CS machinery to the case of two scalar fields.

There is one last subtlety worth pointing out. Here we have used the condition in eq.~(\ref{gtcon}) to derive eq.~(\ref{calrel}). If on the contrary, we wish to view that equation as ``God-given", there is no need to postulate this boundary condition, as it follows directly from the equation itself.

\subsubsection{Second CS equation}

To obtain $\bar{\Gamma}^{(2)}_\theta$ in a finite way we will need to take a second derivative. (Once we have $\bar{\Gamma}^{(2)}_\theta$, we can use the first CS equation to derive $\bar{\Gamma}^{(2)}$.)  To that end, one defines another bare quantity
\be
\Gamma_{\theta\theta}^{(n)}\equiv -i\times\frac{d}{d m_0^2} \Gamma^{(n)}_\theta~, \label{gttrel}
\ee
which is finite for all $n$. There is no need to define a new object like $Z_{\theta\theta}$ in the conversion from $\Gamma_{\theta\theta}^{(n)}$ to its renormalised counterpart. (Better so, as after $\Gamma^{(2)}$, $\Gamma^{(4)}$ and $\Gamma^{(2)}_\theta$ there is no more bare ``divergent" object left whose conversion to its physical counterpart necessitates the introduction of a new object like $Z_{\theta\theta}$.)  Instead, another factor of $Z_\theta$ (compared to eq.~(\ref{relbar2})) suffices:
\be
\Gamma_{\theta\theta}^{(n)} = Z^{n/2}~Z_\theta^2 ~\bar{\Gamma}^{(n)}_{\theta \theta}~. \label{g3t}
\ee

As before, using eqs.~(\ref{relbar2}) and (\ref{g3t}) we can rewrite eq.~(\ref{gttrel}) in terms of physical quantities. The only difference with the previous computation in eqs.~(\ref{defi}) and (\ref{defin}) is the presence of $Z_\theta$ at the right-hand side of eq.~(\ref{gttrel}). This makes that one new quantity enters the scene:
\be
\gamma_\theta \equiv  2 m^2~\left[\frac{\d m^2}{\d m_0^2}\right]^{-1}~\frac{\d \ln{Z_\theta}}{\d m_0^2}~,
\ee
which has to begin at order $\lambda$ as well. With that, it is direct to derive the second equation in eqs.~(\ref{calrel}).

It is worth stressing that by using the CS method in a recursive way, all correlation functions can be found \emph{up to any desired order in the loop expansion}  \cite{Callan}. 
For a rough sketch of the approach, we can again consider the $\lambda \phi^4$ theory. The one loop ``divergent" two- and four-point functions follow from inserting their (finite) tree-level results in the CS equations in eqs.~(\ref{calrel}). In the process of computing these one-loop corrections, one also finds the lowest order contributions to the quantities $\beta$, $\gamma$ and $\gamma_\theta$. With these, the CS procedure can be repeated to derive the next-order corrections to the two- and four-point functions. These yield the next contributions to $\beta$, $\gamma$ and $\gamma_\theta$, and so forth. On the other hand, the correlation functions $\bar{\Gamma}^{(6)}$, $\bar{\Gamma}^{(8)}$ etc. follow from either manifestly convergent diagrams, or from overall convergent diagrams with internal subdivergences. These last ones can be dealt with by the usual skeleton expansion, i.e. by inserting the renormalised two- and four-point functions.

\section{CS method: two scalar fields \label{twof}}

In this Section, we want to generalise the CS method to the case of two scalar fields. We will see that basically, this task comes down to promoting many scalar quantities to two-by-two matrices.

Now we begin from the bare Lagrangian
\be
\L = - \frac{1}{2} \d_\mu \phi_0 \d^\mu \phi_0- \frac{1}{2} \d_\mu \Phi_0 \d^\mu \Phi_0 - \frac{m_0^2}{2} \phi_0^2- \frac{M_0^2}{2} \Phi_0^2 - \frac{\lambda_{\phi,0}}{4!} \phi_0^4 - \frac{\lambda_{\phi\Phi,0}}{4} \phi_0^2\Phi_0^2 - \frac{\lambda_{\Phi,0}}{4!} \Phi_0^4~. 
\ee
We will refer to $\phi$ as the light field and to $\Phi$ as the heavy field (so we will pretend that $m\ll M$), but we will never actually impose that. 

We will see that even if we are only interested in extracting the correlation functions of the light fields, we need to take the ``heavy" correlation functions into account as well. If not, we miss relations that determine the coefficients in the differential equations for the light field's correlation functions.

We need to assume that the three four-point couplings $\lambda_i = \{\lambda_\phi,\lambda_{\phi\Phi}, \lambda_\Phi\}$ are of the same order to be able to make meaningful expansions in these parameters. This order will be generally denoted by $\lambda$.

The bare correlation functions $\Gamma$ are related to the renormalised correlation functions $\bar{\Gamma}$ through
\be
\Gamma^{( n,N )}(m_0,M_0,\lambda_{i,0})  = Z_\phi^{n/2} ~ Z_\Phi^{N/2}~\bar{\Gamma}^{(n,N)}(m,M,\lambda_i)~.\label{brbar}
\ee
Here $n$ denotes the number of ``light" $\phi$ fields, while $N$ denotes the number of ``heavy" $\Phi$ fields.

In leading order, everything is trivial 
\ba
m_0^2 &=& m^2 + \mathcal{O}\left(\lambda\right)~,\qquad \qquad\qquad
\bar{\Gamma}^{(2,0)} = i\left(k^2+m_0^2\right) +\mathcal{O}\left(\lambda\right)~,
\nn\\
M_0^2 &=& M^2 + \mathcal{O}\left(\lambda\right)~,\qquad \qquad \qquad
\bar{\Gamma}^{(0,2)} = i\left(k^2+M_0^2\right) +\mathcal{O}\left(\lambda\right)~,
\nn\\
\lambda_{i,0} &=& \lambda_i + \mathcal{O}\left(\lambda^2\right)~,\qquad \qquad \qquad
\bar{\Gamma}^{(4,0)} = -i\lambda_\phi +\mathcal{O}\left(\lambda^2\right)~,
\nn\\
Z_\phi &=& 1+\mathcal{O}\left(\lambda\right)~,
~~~\qquad \qquad \qquad
\bar{\Gamma}^{(0,4)} = -i\lambda_\Phi +\mathcal{O}\left(\lambda^2\right)~,
\nn\\
Z_\Phi &=& 1+\mathcal{O}\left(\lambda\right)~,
~~~\qquad \qquad \qquad\bar{\Gamma}^{(2,2)} = -i\lambda_{\phi\Phi} +\mathcal{O}\left(\lambda^2\right)~. \label{tree2f}
\ea
The tree level boundary conditions of eq.~(\ref{bountree}) (valid up to all orders) now generalise to
\ba
\bar{\Gamma}^{(2,0)}(0)&=& im^2~,
~~\qquad\qquad \qquad
\bar{\Gamma}^{(4,0)}(0)=-i\lambda_\phi~,
\nn\\
\bar{\Gamma}^{(1,1)}(0)&=& 0~,\qquad\qquad\qquad\qquad
\bar{\Gamma}^{(2,2)}(0) = -i\lambda_{\phi\Phi}~
,\nn\\
\bar{\Gamma}^{(0,2)}(0)&=& iM^2~,
~~\qquad\qquad\qquad
\bar{\Gamma}^{(0,4)}(0)=-i\lambda_\Phi~,
\nn\\
\left[\frac{d}{d k^2}~\bar{\Gamma}^{(0,2)}\right]_{k^2=0}&=& i~,~~\qquad\qquad \qquad \left[\frac{d}{d k^2}~\bar{\Gamma}^{(2,0)}\right]_{k^2=0}= i~.
\label{rcon2}
\ea

\subsection{First CS equation}

In a theory of two fields, we have two kinds of $\theta$-operations: we can take derivatives with respect to the bare small mass or with respect to the bare large mass. So we define
\be
\left(
\begin{array}{ccc}
\Gamma^{(n,N)}_{\theta,m}\\
\Gamma^{(n,N)}_{\theta,M}
\end{array}\right)
\equiv
-i
\left(
\begin{array}{ccc}
\d/\d m_0^2\\
\d/\d M_0^2
\end{array}\right)
\Gamma^{(n,N)}~, \label{2der}
\ee
which are again valid up to all orders in $\lambda$.

All four ``theta-two point functions" are divergent, so they each need a boundary condition. The conditions in eqs.~(\ref{rcon2}) suggest to generalise eq.~(\ref{bounren}) to
\ba
\bar{\Gamma}_{\theta,m}^{(2,0)}(0) &=&1~,
\qquad\qquad\qquad
\bar{\Gamma}_{\theta,m}^{(0,2)}(0) =0~,
\nn\\
\bar{\Gamma}_{\theta,M}^{(2,0)}(0) &=&0~,
\qquad\qquad\qquad
\bar{\Gamma}_{\theta,M}^{(0,2)}(0) =1~.
\label{bgtM2}
\ea
Clearly, these conditions are invariant under simultaneous exchange $m\leftrightarrow M$ and $(2,0)\leftrightarrow (0,2)$.

As the two fields in the theory interact with each other, we now need to generalise the relation between $\Gamma_\theta$ and $\bar{\Gamma}_\theta$ in eq.~(\ref{relbar2}) to a two-by-two matrix equation:
\be
\left(
\begin{array}{ccc}
\Gamma^{(n,N)}_{\theta,m}\\
\Gamma^{(n,N)}_{\theta,M}
\end{array}\right)
=
 Z_\phi^{n/2} ~ Z_\Phi^{N/2}~\underbrace{\left(
\begin{array}{ccc}
Z_{\theta,mm} & Z_{\theta,mM}\\
Z_{\theta,Mm} & Z_{\theta,MM}
\end{array}
\right)}_{\equiv \Z_\theta}
\left(
\begin{array}{ccc}
\bar{\Gamma}^{(n,N)}_{\theta,m}\\
\bar{\Gamma}^{(n,N)}_{\theta,M}
\end{array}\right)~. \label{Misha2}
\ee
To find the two-field equivalent of the first CS equation we begin from the definition in eq.~(\ref{2der}). Using eqs.~(\ref{brbar}) and (\ref{Misha2}) we can again rewrite both sides in terms of renormalised quantities, which gives
\ba
&&i\cdot \Z_\theta\cdot
\left(
\begin{array}{ccc}
\bar{\Gamma}^{(n,N)}_{\theta,m}\\
\bar{\Gamma}^{(n,N)}_{\theta,M}
\end{array}\right)\nn\\
&& \qquad \qquad \qquad
=\Biggl[
 \left(\begin{array}{ccc}
\d m^2/\d m_0^2 \\ \d m^2/\d M_0^2
\end{array}\right) \frac{\d}{\d m^2}+
 \left(\begin{array}{ccc}
\d M^2/\d m_0^2 \\ \d M^2/\d M_0^2
\end{array}\right) \frac{\d}{\d M^2}+
\sum_i \left(\begin{array}{ccc}
\d \lambda_i/\d m_0^2 \\ \d \lambda_i/\d M_0^2
\end{array}\right) \frac{\d}{\d \lambda_i}\nn\\
&& \qquad \qquad \qquad \qquad \qquad \qquad
+\frac{n}{2} \left(\begin{array}{ccc}
\d \ln{Z_\phi}/\d m_0^2 \\ \d\ln{Z_\phi}/\d M_0^2
\end{array}\right) 
+\frac{N}{2} \left(\begin{array}{ccc}
\d \ln{Z_\Phi}/\d m_0^2 \\ \d\ln{Z_\Phi}/\d M_0^2
\end{array}\right) 
\Biggr]\bar{\Gamma}^{(n,N)}~. \label{matreq}
\ea
Here the sum is over $ \lambda_i =\{\lambda_\phi,\lambda_{\phi\Phi}, \lambda_\Phi\}$. At both sides, we have already dropped the factors $Z_\phi^{n/2}$ and $Z_\Phi^{N/2}$. 

Now we would like to avoid equations that contain derivatives with respect to both $m$ and ${M}$. To that end, we express the terms with the mass derivatives through a newly introduced two-by-two ``mass matrix" $\M$:
\ba
&&i\cdot\Z_\theta\cdot
\left(
\begin{array}{ccc}
\bar{\Gamma}^{(n,N)}_{\theta,m}\\
\bar{\Gamma}^{(n,N)}_{\theta,M}
\end{array}\right)
=\Biggl[
\M\cdot 
\left(
\begin{array}{ccc}
\d / \d m^2 \\ \d / \d M^2
\end{array}
\right)
+
\sum_i \left(\begin{array}{ccc}
\d \lambda_i/\d m_0^2 \\ \d \lambda_i/\d M_0^2
\end{array}\right) \frac{\d}{\d \lambda_i}\nn\\
&& \qquad \qquad \qquad \qquad \qquad \qquad
+\frac{n}{2} \left(\begin{array}{ccc}
\d \ln{Z_\phi}/\d m_0^2 \\ \d\ln{Z_\phi}/\d M_0^2
\end{array}\right) 
+\frac{N}{2} \left(\begin{array}{ccc}
\d \ln{Z_\Phi}/\d m_0^2 \\ \d\ln{Z_\Phi}/\d M_0^2
\end{array}\right) 
\Biggr]\bar{\Gamma}^{(n,N)}~,\nn\\
\ea
with $\M$ given by
\be
\M \equiv \left(
\begin{array}{ccc}
\d m^2 /\d m_0^2 & \d M^2 /\d m_0^2 \\
\d m^2 /\d M_0^2 & \d M^2 /\d M_0^2 
\end{array}
\right)~.
\ee
Multiplying (from the left) our equation by the inverse of the mass matrix, just like we did in the one-field case, gives
\ba
&&i\cdot
\M
^{-1}
\cdot \Z_\theta\cdot
\left(
\begin{array}{ccc}
\bar{\Gamma}^{(n,N)}_{\theta,m}\\
\bar{\Gamma}^{(n,N)}_{\theta,M}
\end{array}\right)
=\Biggl[
\left(
\begin{array}{ccc}
\d / \d m^2 \\ \d / \d M^2
\end{array}
\right)
+
\sum_i 
\M^{-1}\cdot
\left(\begin{array}{ccc}
\d \lambda_i/\d m_0^2 \\ \d \lambda_i/\d M_0^2
\end{array}\right) \frac{\d}{\d \lambda_i}\nn\\
&& \qquad \qquad \qquad \qquad \qquad \qquad \qquad
+\frac{n}{2}\cdot
\M^{-1}\cdot
 \left(\begin{array}{ccc}
\d \ln{Z_\phi}/\d m_0^2 \\ \d\ln{Z_\phi}/\d M_0^2
\end{array}\right)\nn\\
&& \qquad \qquad \qquad \qquad \qquad \qquad \qquad
+\frac{N}{2}\cdot
\M^{-1}\cdot
 \left(\begin{array}{ccc}
\d \ln{Z_\Phi}/\d m_0^2 \\ \d\ln{Z_\Phi}/\d M_0^2
\end{array}\right) 
\Biggr]\bar{\Gamma}^{(n,N)}~.
\ea
At this point, we are ready to absorb all bare quantities in new definitions. The two-field equivalent of eqs.~(\ref{3def}) becomes
\ba
\G &\equiv & \M^{-1} \cdot \Z_\theta~,\nn\\
\left(\begin{array}{ccc}
\frac{1}{2m^2} \beta_{\lambda_{i,m}} \\ \frac{1}{2M^2} \beta_{\lambda_{i,M}}
\end{array}\right) 
&\equiv &
\M^{-1}
\left(\begin{array}{ccc}
\d \lambda_i/\d m_0^2 \\ \d \lambda_i/\d M_0^2
\end{array}\right)~,\nn\\
&&\nn\\
 \left(\begin{array}{ccc}
\frac{1}{m^2} \gamma_{\phi,m} \\ \frac{1}{M^2} \gamma_{\phi,M}
\end{array}\right) 
&\equiv&
\M^{-1}
 \left(\begin{array}{ccc}
\d \ln{Z_\phi}/\d m_0^2 \\ \d\ln{Z_\phi}/\d M_0^2
\end{array}\right)~, \nn\\
 \left(\begin{array}{ccc}
\frac{1}{m^2} \gamma_{\Phi,m} \\ \frac{1}{M^2} \gamma_{\Phi,M}
\end{array}\right) 
&\equiv&
\M^{-1}
 \left(\begin{array}{ccc}
\d \ln{Z_\Phi}/\d m_0^2 \\ \d\ln{Z_\Phi}/\d M_0^2
\end{array}\right)~. \label{nutrel}
\ea
Again, from the tree level results in eq.~(\ref{tree2f}), we find that all $\beta$'s begin at order $\lambda^2$ while all $\gamma$'s begin at order $\lambda$. Now we are at
\ba
&&i\cdot
\G\cdot
\left(
\begin{array}{ccc}
\bar{\Gamma}^{(n,N)}_{\theta,m}\\
\bar{\Gamma}^{(n,N)}_{\theta,M}
\end{array}\right)
=\Biggl[
\left(
\begin{array}{ccc}
\d / \d m^2 \\ \d / \d M^2
\end{array}
\right)
+
\sum_i 
\left(\begin{array}{ccc}
\frac{1}{2m^2} \beta_{\lambda_{i,m}} \\ \frac{1}{2M^2} \beta_{\lambda_{i,M}}
\end{array}\right) \frac{\d}{\d \lambda_i}\nn\\
&& \qquad \qquad \qquad \qquad \qquad 
+\frac{n}{2}
 \left(\begin{array}{ccc}
\frac{1}{m^2} \gamma_{\phi,m} \\ \frac{1}{M^2} \gamma_{\phi,M}
\end{array}\right)
+\frac{N}{2}
 \left(\begin{array}{ccc}
\frac{1}{m^2} \gamma_{\Phi,m} \\ \frac{1}{M^2} \gamma_{\Phi,M}
\end{array}\right) 
\Biggr]\bar{\Gamma}^{(n,N)}~. \label{bijnakl}
\ea
Next, we use the renormalisation conditions on $\bar{\Gamma}^{(2,0)}$ and $\bar{\Gamma}^{(0,2)}$ to determine the unknown matrix $\G$. We find the two-field equivalent of eq.~(\ref{remov}):
\be
\G = \left(
\begin{array}{ccc}
1+\gamma_{\phi,m} &  \frac{M^2}{m^2} \gamma_{\Phi,m} \\
 \frac{m^2}{M^2} \gamma_{\phi,M} & 1+\gamma_{\Phi,M}
\end{array}
\right)~,
 \label{shandig}
\ee
up to all orders in the ``combined coupling constant" $\lambda$.

For clarity, we can separate the two equations contained in the matrix equation (\ref{bijnakl}). The upper one gives, after multiplying by $2m^2$,
\ba
2i\left[m^2\left(1+\gamma_{\phi,m}\right)\bar{\Gamma}^{(n,N)}_{\theta,m} + M^2 \gamma_{\Phi,m}\bar{\Gamma}^{(n,N)}_{\theta,M}\right] &=& \Biggl[2m^2\frac{\d}{\d m^2}+\sum_i\beta_{\lambda_i,m}\frac{\d}{\d \lambda_i}\nn\\
&&   \qquad\qquad+n\cdot \gamma_{\phi,m}+N\cdot \gamma_{\Phi,m}\Bigg]\bar{\Gamma}^{(n,N)}  ~. \nn\\\label{clight}
\ea
As a consistency check, we note that omitting the last term at the left-hand side and the last term at the right-hand side, and taking the ``sum" over just $\lambda$, brings us back to the first equation in eq.~(\ref{calrel}).

For the lower one, we get, after multiplying by $2M^2$,
\ba
2i\left[M^2\left(1+\gamma_{\Phi,M}\right)\bar{\Gamma}^{(n,N)}_{\theta,M} + m^2 \gamma_{\phi,M}\bar{\Gamma}^{(n,N)}_{\theta,m}\right] &=& \Biggl[2M^2\frac{\d}{\d M^2}+\sum_i\beta_{\lambda_i,M}\frac{\d}{\d \lambda_i}\nn\\
&&   \qquad\qquad+n\cdot \gamma_{\phi,M}+N\cdot \gamma_{\Phi,M}\Biggr]\bar{\Gamma}^{(n,N)}~.\nn\\\label{cheavy}
\ea
With these two equations in hand, we can again solve for derivatives with respect to both $m^2$ and $M^2$ of all renormalised correlation functions that traditionally require divergent Feynman diagrams: $\bar{\Gamma}^{(2,0)}$, $\bar{\Gamma}^{(0,2)}$, $\bar{\Gamma}^{(4,0)}$, $\bar{\Gamma}^{(0,4)}$ and $\bar{\Gamma}^{(2,2)}$. In passing by, these ten equations precisely yield expressions for the six $\beta$'s and four $\gamma$'s defined in eqs.~(\ref{nutrel}). We find that for every coupling $\lambda_i$, the sum of $\beta_{\lambda_i,m}$ and $\beta_{\lambda_i,M}$ yields its beta-function. The same goes for the wave-function renormalisations $\gamma_\phi$ and $\gamma_\Phi$. 

Just like in the one field case, these ten expressions help to begin the computation of the next order coefficients of these correlation functions, which yields the next order coefficients to the $\beta$'s and $\gamma$'s, etcetera. 

Clearly, to generalise from two fields to $n$ fields one should just promote all two-by-two matrices in this section to $n$-by-$n$ matrices.

\subsection{Second CS equation}

The last divergent elements that we need to compute to make the above program consistent are the four correlation functions $\Gamma^{(2,0)}_{\theta,m}$, $\Gamma^{(2,0)}_{\theta,M}$, $\Gamma^{(0,2)}_{\theta,m}$ and $\Gamma^{(0,2)}_{\theta,M}$. In order to do so, we define a second $\theta$-operation through, in matrix notation already,
\be
i\cdot \left(
\begin{array}{ccc}
\Gamma_{\theta\theta,mm} & \Gamma_{\theta\theta,mM}\\
\Gamma_{\theta \theta, Mm}&\Gamma_{\theta \theta,MM}
\end{array} 
\right)
=
\left(
\begin{array}{ccc}
\d \Gamma_{\theta,m}/\d m_0^2 & \d \Gamma_{\theta,M}/\d m_0^2 \\ 
\d \Gamma_{\theta,m}/\d M_0^2 & \d \Gamma_{\theta,M}/\d M_0^2
\end{array}
\right)~.  \label{cal2eq}
\ee
By construction, the two off-diagonal objects should be equal to each other: partial derivatives commute.

To rewrite the left-hand side of this equation in terms of renormalised quantities we follow the same philosophy as in eq.~(\ref{Misha2}), which gives
\be
\left(
\begin{array}{ccc}
\Gamma_{\theta\theta,mm} & \Gamma_{\theta\theta,mM}\\
\Gamma_{\theta \theta, Mm}&\Gamma_{\theta \theta,MM}
\end{array}
\right)
= 
 Z_\phi^{n/2} ~ Z_\Phi^{N/2}~\mathcal{Z}_\theta \times
\left(
\begin{array}{ccc}
\bar{\Gamma}_{\theta\theta,mm} & \bar{\Gamma}_{\theta\theta,mM}\\
\bar{\Gamma}_{\theta \theta, Mm}&\bar{\Gamma}_{\theta \theta,MM}
\end{array}
\right)
\times \left(\mathcal{Z}_\theta\right)^T~. \label{barth}
\ee
Rewriting the right-hand side of eq.~(\ref{cal2eq}) gives, after taking derivatives
\ba
&&\left(
\begin{array}{ccc}
\d \Gamma_{\theta,m}/\d m_0^2 & \d \Gamma_{\theta,M}/\d m_0^2 \\ 
\d \Gamma_{\theta,m}/\d M_0^2 & \d \Gamma_{\theta,M}/\d M_0^2
\end{array}
\right)\nn\\
&&\qquad =Z_\phi^{n/2}~Z_\Phi^{N/2}~\M\times \Biggl(
\Biggl[
\left(\begin{array}{ccc}
\d / \d m^2 \\ \d/\d M^2\end{array}\right)
+\sum_i \left(\begin{array}{ccc}
\beta_{\lambda_i,m}/2m^2 \\ \beta_{\lambda_i,M}/2M^2
\end{array}\right)
\frac{\d}{\d \lambda_i}\nn\\
&& \qquad \qquad \qquad\qquad\qquad\qquad
+\frac{n}{2}\left(\begin{array}{ccc}
\gamma_{\phi,m}/m^2 \\ \gamma_{\phi,M}/M^2
\end{array}\right)
+\frac{N}{2}\left(\begin{array}{ccc}
\gamma_{\Phi,m}/m^2 \\ \gamma_{\Phi,M}/M^2
\end{array}\right)
\Biggr]
\left(\begin{array}{ccc}
\bar{\Gamma}_{\theta,m} & \bar{\Gamma}_{\theta,M}\end{array}\right)\nn\\
&& \qquad \qquad\qquad \qquad \qquad\qquad \qquad+\left[ 
\frac{\gamma_{\theta,m}}{2m^2} ~\bar{\Gamma}_{\theta,m}
+
\frac{\gamma_{\theta,M}}{2M^2} ~\bar{\Gamma}_{\theta,M}
\right]
\Biggr)
\left(\Z_\theta\right)^T~.
\ea
In the last line we now get mixing between $\bar{\Gamma}_{\theta,m}$ and $\bar{\Gamma}_{\theta,M}$. The two new two-by-two matrices $\gamma_{\theta,m}$ and $\gamma_{\theta,M}$ contain eight new quantities in total:
\ba
\frac{\gamma_{\theta,m}}{2m^2} &\equiv &
 \M^{-1}\cdot \left(\begin{array}{ccc}
\d Z_{\theta,mm}/ \d m_0^2 & \d Z_{\theta,Mm}/ \d m_0^2 \\
\d Z_{\theta,mm}/ \d M_0^2 & \d Z_{\theta,Mm}/ \d M_0^2
\end{array}\right)
\cdot
\left(\Z_\theta^T\right)^{-1} ~,\nn\\
&\equiv &
 \M^{-1}\cdot \left(\begin{array}{ccc}
\gamma_{\theta,mmm} & \gamma_{\theta,Mmm} \\
\gamma_{\theta,mmM} & \gamma_{\theta,MmM}
\end{array}\right) \cdot
\left(\Z_\theta^T\right)^{-1} ~, \nn\\
&&\nn\\
\frac{\gamma_{\theta,M}}{2M^2} &\equiv &
 \M^{-1}\cdot \left(\begin{array}{ccc}
\d Z_{\theta,mM}/ \d m_0^2  & \d Z_{\theta,MM}/ \d m_0^2\\
\d Z_{\theta,mM}/ \d M_0^2& \d Z_{\theta,MM}/ \d M_0^2
\end{array}\right)
\cdot
\left(\Z_\theta^T\right)^{-1}~,\nn\\
&\equiv &
  \M^{-1}\cdot \left(\begin{array}{ccc}
\gamma_{\theta,mMm} & \gamma_{\theta,MMm} \\
\gamma_{\theta,mMM} & \gamma_{\theta,MMM}
\end{array}\right) \cdot
\left(\Z_\theta^T\right)^{-1}~.
\ea
Since we are going to end up with 8 equations (derivatives with respect to both $m^2$ and $M^2$ of the four $\bar{\Gamma}_\theta$'s that we are after), applying the renormalisation conditions will precisely yield 8 equations for these 8 new quantities.

Finally, setting eq.~(\ref{barth}) equal to this last equation gives, upon using eq.~(\ref{shandig}), the two-by-two matrix equation
\ba
 i\cdot &&\left(
\begin{array}{ccc}
1+\gamma_{\phi,m} &  \frac{M^2}{m^2} \gamma_{\Phi,m} \\
 \frac{m^2}{M^2} \gamma_{\phi,M} & 1+\gamma_{\Phi,M}
\end{array}
\right) \times  \left(
\begin{array}{ccc}
\bar{\Gamma}_{\theta\theta,mm} & \bar{\Gamma}_{\theta\theta,mM}\\
\bar{\Gamma}_{\theta \theta, Mm}&\bar{\Gamma}_{\theta \theta,MM}
\end{array}
\right)\nn\\
&& \qquad \qquad=
\Biggl[
\left(\begin{array}{ccc}
\d / \d m^2 \\ \d/\d M^2\end{array}\right)
+\sum_i \left(\begin{array}{ccc}
\beta_{\lambda_i,m}/2m^2 \\ \beta_{\lambda_i,M}/2M^2
\end{array}\right)
\frac{\d}{\d \lambda_i}\nn\\
&& \qquad \qquad \qquad\qquad
+\frac{n}{2}\left(\begin{array}{ccc}
\gamma_{\phi,m}/m^2 \\ \gamma_{\phi,M}/M^2
\end{array}\right)
+\frac{N}{2}\left(\begin{array}{ccc}
\gamma_{\Phi,m}/m^2 \\ \gamma_{\Phi,M}/M^2
\end{array}\right)
\Biggr]
\left(\begin{array}{ccc}
\bar{\Gamma}_{\theta,m} & \bar{\Gamma}_{\theta,M}\end{array}\right)\nn\\
&& \qquad \qquad\qquad+
\frac{1}{2m^2} 
\times
 \left(\begin{array}{ccc}
\gamma_{\theta,mmm} & \gamma_{\theta,Mmm} \\
\gamma_{\theta,mmM} & \gamma_{\theta,MmM}
\end{array}\right) 
 ~\bar{\Gamma}_{\theta,m}\nn\\
 &&\qquad\qquad \qquad
+
\frac{1}{2M^2}
\times
\left(\begin{array}{ccc}
\gamma_{\theta,mMm} & \gamma_{\theta,MMm} \\
\gamma_{\theta,mMM} & \gamma_{\theta,MMM}
\end{array}\right)
 ~\bar{\Gamma}_{\theta,M}~, \label{matr2}
\ea
which is again entirely made of finite quantities. Each of the four component equations in this matrix equation can be evaluated for $(n=2,N=0)$ and $(n=0,N=2)$, which leaves us with the desired eight equations.

\section{Computations}

We will use this Section to explicitly show the computation of the first corrections to the two- and four-point correlation functions of the light field $\phi$. The computations of all other correlation functions proceed along precisely the same lines.

\subsection{Four-point function $\bar{\Gamma}^{(4,0)}$}

\begin{figure}[!h]
\begin{center}
\begin{tikzpicture}
[line width=1.5 pt, scale=1.5]

\begin{scope}
\node at (0,0) {$\bar{\Gamma}^{(4,0)}$};
\end{scope}

\begin{scope}
\node at (0.5,0) {$=$};
\end{scope}

\begin{scope}
[shift={(1,0)}]
\draw (-0.25,0.25)--(0.25,-0.25);
\draw (-0.25,-0.25)--(0.25,0.25);
\draw [fill=black](0,0) circle (0.05cm);
\node at (0.675,0) {$+$};
\end{scope}

\begin{scope}
[shift={(2,0)}]
\draw (0,0.25)--(0.25,0);
\draw (0,-0.25)--(0.25,0);
\draw [fill=black](0.25,0) circle (0.05cm);
\draw (0.5,0) circle (0.25cm);
\draw [fill=black](0.75,0) circle (0.05cm);
\draw (0.75,0)--(1,0.25);
\draw (0.75,0)--(1,-0.25);
\node at (1.5,0) {$+$};
\end{scope}

\begin{scope}
[shift={(4,0)}]
\draw (0,0.25)--(0.25,0);
\draw (0,-0.25)--(0.25,0);
\draw [fill=black](0.25,0) circle (0.05cm);
\draw [dashed](0.5,0) circle (0.25cm);
\draw [fill=black](0.75,0) circle (0.05cm);
\draw (0.75,0)--(1,0.25);
\draw (0.75,0)--(1,-0.25);
\node at (2,0) {$+~~\mathcal{O}\left(\lambda^3\right)$};
\end{scope}

\begin{scope}
[shift={(0,-1)}]
\node at (0,0) {$\bar{\Gamma}^{(4,0)}_{\theta,m}$};
\end{scope}

\begin{scope}
[shift={(0,-1)}]
\node at (0.5,0) {$=$};
\end{scope}

\begin{scope}
[shift={(2,-1)}]
\draw (0,0.25)--(0.25,0);
\draw (0,-0.25)--(0.25,0);
\draw [fill=black](0.25,0) circle (0.05cm);
\draw (0.5,0) circle (0.25cm);
\draw (0.4,0.15)--(0.6,0.35);
\draw (0.4,0.35)--(0.6,0.15);
\draw [fill=black](0.75,0) circle (0.05cm);
\draw (0.75,0)--(1,0.25);
\draw (0.75,0)--(1,-0.25);
\node at (-0.25,0){$2~\times$};
\node at (4,0) {$+~~~\mathcal{O}\left(\lambda^3\right)$};
\end{scope}

\begin{scope}
[shift={(4,-2)}]
\draw (0,0.25)--(0.25,0);
\draw (0,-0.25)--(0.25,0);
\draw [fill=black](0.25,0) circle (0.05cm);
\draw [dashed] (0.5,0) circle (0.25cm);
\draw (0.4,0.15)--(0.6,0.35);
\draw (0.4,0.35)--(0.6,0.15);
\draw [fill=black](0.75,0) circle (0.05cm);
\draw (0.75,0)--(1,0.25);
\draw (0.75,0)--(1,-0.25);
\node at (-0.25,0){$2~\times$};
\node at (2,0) {$+~~~\mathcal{O}\left(\lambda^3\right)$};
\end{scope}

\begin{scope}
[shift={(0,-2)}]
\node at (0,0) {$\bar{\Gamma}^{(4,0)}_{\theta,M}$};
\end{scope}

\begin{scope}
[shift={(0,-2)}]
\node at (0.5,0) {$=$};
\end{scope}

\end{tikzpicture}
\caption{
Graphical depiction of the
 $(\theta,m)$ and $(\theta,M)$ operations on the one-loop corrections to the light field's four-point function. The continuous line denotes the light field $\phi$, the dashed line denotes the heavy field $\Phi$.
}
\end{center}
\end{figure}

 From the tree-level analysis, we know already that
\be
\bar{\Gamma}^{(4,0)} = -i\lambda_\phi+\mathcal{O}\left(\lambda^2\right)~,
\ee
and the goal now is to find the order $\lambda^2$ contribution. In order to do so, we have the boundary condition
\be
\bar{\Gamma}^{(4,0)}(0) = -i\lambda_\phi~,
\ee
and the readily computed convergent quantities
\ba
\bar{\Gamma}_{\theta,m}^{(4,0)} &=& -\frac{\lambda_\phi^2}{32\pi^2}\sum_{\rm 3~opt}~\int_0^1 dx~ \frac{1} {x(1-x)\kappa_1^2+m^2}+\mathcal{O}\left(\lambda^3\right)~, \label{gtm}\\
\bar{\Gamma}_{\theta,M}^{(4,0)} &=& -\frac{\lambda_{\phi\Phi}^2}{32\pi^2}\sum_{\rm 3~opt}~\int_0^1 dx~ \frac{1} {x(1-x)\kappa_1^2+M^2}+\mathcal{O}\left(\lambda^3\right)~. 
\ea

For $n=4, N=0$ the light and heavy versions of the first CS equation that we derived in eqs.~(\ref{clight}) and (\ref{cheavy}) look like
\ba
2i\left[m^2\left(1+\gamma_{\phi,m}\right)\bar{\Gamma}^{(4,0)}_{\theta,m} + M^2 \gamma_{\Phi,m}\bar{\Gamma}^{(4,0)}_{\theta,M}\right] &=& \left[2m^2\frac{\d}{\d m^2}+\sum_i\beta_{\lambda_i,m}\frac{\d}{\d \lambda_i}+4\cdot \gamma_{\phi,m}\right]\bar{\Gamma}^{(4,0)} ~,\nn\\
2i\left[M^2\left(1+\gamma_{\Phi,M}\right)\bar{\Gamma}^{(4,0)}_{\theta,M} + m^2 \gamma_{\phi,M}\bar{\Gamma}^{(4,0)}_{\theta,m}\right] &=& \left[2M^2\frac{\d}{\d M^2}+\sum_i\beta_{\lambda_i,M}\frac{\d}{\d \lambda_i}+4\cdot \gamma_{\phi,M}\right]\bar{\Gamma}^{(4,0)}~.\nn\\
\ea
First, we look at the light equation. We know that all $\gamma$'s begin at order $\lambda$, and all $\beta$'s at order $\lambda^2$. Therefore the left-hand side begins at order $\lambda^2$. The right-hand side begins at order $\lambda^2$ as well. We have for the order $\lambda^2$ contribution to this equation
\be
2i m^2\cdot \left[\bar{\Gamma}^{(4,0)}_{\theta,m}\right]_{\lambda^2} = 2m^2\frac{\d}{\d m^2}\left[\bar{\Gamma}^{(4,0)}\right]_{\lambda^2} +\left[\beta_{\lambda_\phi,m}\right]_{\lambda^2}\cdot -i +4 \left[\gamma_{\phi,m}\right]_\lambda \cdot -i\lambda_\phi~. \label{2im2}
\ee
(Again, the notation $\left[\bar{\Gamma}^{(4,0)}_{\theta,m}\right]_{\lambda^2} $ denotes the order $\lambda^2$ contribution to $\bar{\Gamma}^{(4,0)}_{\theta,m}$.)
Evaluating this equation at $k^2=0$ we can use the renormalisation condition $\bar{\Gamma}^{(4,0)}(0) = -i\lambda_\phi$ to end up with
\be
\frac{3 \lambda_\phi^2}{16\pi^2} = \left[\beta_{\lambda_\phi,m}\right]_{\lambda^2}+4\left[\gamma_{\phi,m}\right]_\lambda \cdot \lambda_\phi~. \label{us1a}
\ee
With this relation, our equation in eq.~(\ref{2im2}) can be rewritten as
\be
 \frac{\d}{\d m^2}\left[\bar{\Gamma}^{(4,0)}\right]_{\lambda^2} = -\frac{i\lambda_\phi^2}{32\pi^2}\sum_{\rm 3~opt}~\int_0^1 dx~ \frac{1} {x(1-x)\kappa_1^2+m^2} +\frac{3i\lambda_\phi^2}{32\pi^2}\cdot \frac{1}{m^2}~.
\ee
Integrating gives
\be
\left[\bar{\Gamma}^{(4,0)}\right]_{\lambda^2}= \frac{i\lambda_\phi^2}{32\pi^2}\sum_{\rm 3~opt}~\int_0^1 dx~ \ln{\frac{m^2}{x(1-x)\kappa_1^2+m^2}}+c_1~, \label{eQQQ}
\ee
where we have already split off a term proportional to $\ln{(m^2)}$ from the integration constant to get something that resembles the one-field solution. $c_1$ is a dimensionless integration constant, which by construction does not depend on $m$.

Now we look at the heavy equation. At leading order (i.e., order $\lambda^2$) we have
\be
2i M^2  ~\left[\bar{\Gamma}^{(4,0)}_{\theta,M}\right]_{\lambda^2}=  2M^2 ~\frac{\d}{\d M^2}\left[\bar{\Gamma}^{(4,0)}\right]_{\lambda^2}+\left[\beta_{\lambda_\phi,M}\right]_{\lambda^2}\cdot -i+
4\left[\gamma_{\phi,M}\right]_\lambda\cdot -i\lambda_\phi~.
\ee
Evaluating the order $\lambda^2$ contribution to this equation at $k^2=0$ now gives the condition
\be
\frac{3 \lambda_{\phi\Phi}^2}{16\pi^2} = \left[\beta_{\lambda_\phi,M}\right]_{\lambda^2}+4\left[\gamma_{\phi,M}\right]_\lambda\cdot \lambda_\phi~. \label{us2}
\ee
With this relation, the order $\lambda^2$ contribution to our equation  can be rewritten as
\be
 \frac{\d}{\d M^2}\left[\bar{\Gamma}^{(4,0)}\right]_{\lambda^2} = -\frac{i\lambda_{\phi\Phi}^2}{32\pi^2}\sum_{\rm 3~opt}~\int_0^1 dx~ \frac{1} {x(1-x)\kappa_1^2+M^2} +\frac{3i\lambda_{\phi\Phi}^2}{32\pi^2}\cdot \frac{1}{M^2}~.
\ee
Integrating gives
\be
\left[\bar{\Gamma}^{(4,0)}\right]_{\lambda^2}= \frac{i\lambda_{\phi\Phi}^2}{32\pi^2}\sum_{\rm 3~opt}~\int_0^1 dx~ \ln{\frac{M^2}{x(1-x)\kappa_1^2+M^2}}+c_2~,
\ee
with $c_2$ a new dimensionless integration constant that cannot depend on $M$.

Combining the first differential equation in eq.~(\ref{eQQQ}) with this one, the only solution is the conventional result
\ba
\bar{\Gamma}^{(4,0)}(\kappa_1)&=& -i\lambda_\phi + \sum_{\rm 3~ opt}\frac{i\lambda_\phi^2}{32\pi^2}~\int_0^1 dx~\ln{\frac{m^2}{x(1-x)\kappa_1^2+m^2}}
\nn\\
&& \qquad \qquad 
+\sum_{\rm 3~ opt}\frac{i\lambda_{\phi\Phi}^2}{32\pi^2}~\int_0^1 dx~\ln{\frac{M^2}{x(1-x)\kappa_1^2+M^2}}+ \mathcal{O}\left(\lambda^3\right)~.\label{res1}
\ea

\subsection{Two-point function }

 \begin{figure}[!h]
\begin{center}
\begin{tikzpicture}
[line width=1.5 pt, scale=1.5]

\begin{scope}
\node at (0,0) {$\bar{\Gamma}^{(2,0)}$};
\end{scope}

\begin{scope}
\node at (0.5,0) {$=$};
\end{scope}

\begin{scope}
[shift={(1,0)}]
\draw (0,0)--(0.75,0);
\node at (1.125,0) {$-$};
\end{scope}

\begin{scope}
[shift={(3,0)}]
\draw (0,0)--(1,0);
\draw [fill=black](0.5,0) circle (0.05cm);
\draw (0.5,0.25) circle (0.25cm);
\node at (1.5,0) {$-$};
\end{scope}

\begin{scope}
[shift={(5,0)}]
\draw (0,0)--(1,0);
\draw [fill=black](0.5,0) circle (0.05cm);
\draw [dashed] (0.5,0.25) circle (0.25cm);
\node at (2,0) {$+~~\O\left(\lambda^2\right)$};
\end{scope}

\begin{scope}
[shift={(0,-1)}]
\node at (0,0) {$\bar{\Gamma}^{(2,0)}_{\theta,m}$};
\end{scope}

\begin{scope}
[shift={(0,-1)}]
\node at (0.5,0) {$=$};
\end{scope}

\begin{scope}
[shift={(1,-1)}]
\node at (0.375,0) {$1$};
\node at (1.125,0) {$-$};
\end{scope}

\begin{scope}
[shift={(3,-1)}]
\draw (0,0)--(1,0);
\draw [fill=black](0.5,0) circle (0.05cm);
\draw (0.4,0.4)--(0.6,0.6);
\draw (0.4,0.6)--(0.6,0.4);
\draw (0.5,0.25) circle (0.25cm);
\node at (4,0) {$+~~\O\left(\lambda^2\right)$};
\end{scope}

\begin{scope}
[shift={(0,-2)}]
\node at (0,0) {$\bar{\Gamma}^{(2,0)}_{\theta,M}$};
\end{scope}

\begin{scope}
[shift={(0,-2)}]
\node at (0.5,0) {$=$};
\end{scope}

\begin{scope}
[shift={(5,-2)}]
\node at (-0.5,0) {$-$};
\draw (0,0)--(1,0);
\draw [fill=black](0.5,0) circle (0.05cm);
\draw(0.4,0.4)--(0.6,0.6);
\draw(0.4,0.6)--(0.6,0.4);

\draw [dashed] (0.5,0.25) circle (0.25cm);
\node at (2,0) {$+~~\O\left(\lambda^2\right)$};
\end{scope}

\begin{scope}
[shift={(0,-3)}]
\node at (0,0) {$\bar{\Gamma}^{(2,0)}_{\theta\theta,mm}$};
\end{scope}

\begin{scope}
[shift={(0,-3)}]
\node at (0.5,0) {$=$};
\end{scope}

\begin{scope}
[shift={(3,-3)}]
\draw (0,0)--(1,0);
\node at (-0.35,0) {$-2~\times$};
\draw [fill=black](0.5,0) circle (0.05cm);
\draw (0.4,0.4)--(0.6,0.6);
\draw (0.4,0.6)--(0.6,0.4);
\draw (0.15,0.15)--(0.35,0.35);
\draw (0.35,0.15)--(0.15,0.35);
\draw (0.5,0.25) circle (0.25cm);
\node at (4,0) {$+~~\O\left(\lambda^2\right)$};
\end{scope}

\begin{scope}
[shift={(0,-4)}]
\node at (0,0) {$\bar{\Gamma}^{(2,0)}_{\theta\theta,MM}$};
\end{scope}

\begin{scope}
[shift={(0,-4)}]
\node at (0.5,0) {$=$};
\end{scope}

\begin{scope}
[shift={(5,-4)}]
\draw (0,0)--(1,0);
\node at (-0.35,0) {$-2~\times$};
\draw [fill=black](0.5,0) circle (0.05cm);
\draw (0.4,0.4)--(0.6,0.6);
\draw (0.4,0.6)--(0.6,0.4);
\draw (0.15,0.15)--(0.35,0.35);
\draw (0.35,0.15)--(0.15,0.35);
\draw[dashed] (0.5,0.25) circle (0.25cm);
\node at (2,0) {$+~~\O\left(\lambda^2\right)$};
\end{scope}

\begin{scope}
[shift={(0,-5)}]
\node at (0,0) {$\bar{\Gamma}^{(2,0)}_{\theta\theta,mM}$};
\node at (0.5,0) {$=$};
\node at (7,0) {$~~\O\left(\lambda^2\right)$};
\end{scope}

\end{tikzpicture}

\caption{ Graphical depiction of the $(\theta,m)$ and $(\theta,M)$ operations on the one-loop correction to the light field's two-point function. The continuous line denotes the light field $\phi$, the dashed line denotes the heavy field $\Phi$. The minus signs are caused by the fact that in our adopted sign convention, the two-point function is given by the tree-level term minus the contributions from all correcting Feynman diagrams. \label{2th}
}

\end{center}
\end{figure}

\subsubsection{$\bar{\Gamma}^{(2,0)}_{\theta,m}$ and $\bar{\Gamma}^{(2,0)}_{\theta,M}$}

Our first goal is to find the order $\lambda$ contributions to $\bar{\Gamma}^{(2,0)}_{\theta,m}$ and $\bar{\Gamma}^{(2,0)}_{\theta,M}$. The necessary renormalisation conditions are, up to all orders,
\be
\bar{\Gamma}^{(2,0)}_{\theta,m}(0) = 1~,\qquad \qquad \bar{\Gamma}^{(2,0)}_{\theta,M}=0~.
\ee
Three convergent computations lead to the beginning points
\ba
\bar{\Gamma}^{(2,0)}_{\theta\theta,mm}&=& -\frac{i \lambda_{\phi}}{32\pi^2}\cdot \frac{1}{m^2}+\mathcal{O}\left(\lambda^2\right)~,\nn\\
\bar{\Gamma}^{(2,0)}_{\theta\theta,mM}&=&\bar{\Gamma}^{(2,0)}_{\theta\theta,Mm}=\mathcal{O}\left(\lambda^2\right)~,\nn\\
\bar{\Gamma}^{(2,0)}_{\theta\theta,MM}&=& -\frac{i \lambda_{\phi\Phi}}{32\pi^2}\cdot \frac{1}{M^2}+\mathcal{O}\left(\lambda^2\right)~.
\ea
Now, at order $\lambda^0$ we have
\ba
\bar{\Gamma}^{(2,0)}_{\theta,m} &=& 1+\mathcal{O}\left(\lambda\right)~,\nn\\
\bar{\Gamma}^{(2,0)}_{\theta,M} &=& \mathcal{O}\left(\lambda\right)~.
\ea

Armed with these boundary conditions, we can turn to Callan's second equation that we derived in eq.~(\ref{matr2}). Being a two-by-two matrix equation, there are four equations to be extracted. We multiply the upper two by $2m^2$ and the lower two by $2M^2$.  Evaluating the order $\lambda$ contributions to these four equations then gives, respectively,
\ba
2im^2\cdot  -\frac{i \lambda_{\phi}}{32\pi^2}\cdot \frac{1}{m^2} &=& 2m^2 \frac{\d}{\d m^2}\left[\bar{\Gamma}^{(2,0)}_{\theta,m}\right]_\lambda + \left(2\left[\gamma_{\phi,m}\right]_\lambda+\left[\gamma_{\theta,mmm}\right]_\lambda\right)\cdot 1\nn\\
0&=& 2M^2 \frac{\d}{\d M^2}\left[\bar{\Gamma}^{(2,0)}_{\theta,m}\right]_\lambda + \left(2\left[\gamma_{\phi,M}\right]_\lambda+\frac{M^2}{m^2}\left[\gamma_{\theta,mmM}\right]_\lambda\right)\cdot 1\nn\\
0&=& 2m^2\frac{\d}{\d m^2}\left[\bar{\Gamma}^{(2,0)}_{\theta,M}\right]_\lambda + \left[\gamma_{\theta,Mmm}\right]_\lambda \cdot 1\nn\\
2iM^2 \cdot -\frac{i \lambda_{\phi\Phi}}{32\pi^2}\cdot \frac{1}{M^2} &=&  2M^2\frac{\d}{\d M^2}\left[\bar{\Gamma}^{(2,0)}_{\theta,M}\right]_\lambda +\frac{M^2}{m^2}\left[\gamma_{\theta,MmM}\right]_\lambda \cdot 1~. \label{4eq}
\ea
We anticipate already that in the next subsection, the first CS equation will yield
\be
\left[\gamma_{\phi,m}\right]_\lambda = \left[\gamma_{\phi,M}\right]_\lambda =0~.
\ee
With that, evaluating eqs.~(\ref{4eq}) at $k^2=0$ gives
\ba
 \frac{ \lambda_{\phi}}{16\pi^2} &=& \left[\gamma_{\theta,mmm}\right]_\lambda\nn\\
0&=& \frac{M^2}{m^2}\left[\gamma_{\theta,mmM}\right]_\lambda\nn\\
0&=&  \left[\gamma_{\theta,Mmm}\right]_\lambda \nn\\
 \frac{ \lambda_{\phi\Phi}}{16\pi^2}&=&  \frac{M^2}{m^2}\left[\gamma_{\theta,MmM}\right]_\lambda~. 
\ea
As a consistency check, we see that this is indeed a straightforward generalisation of the one-field result
\be
\gamma_\theta = \frac{\lambda}{16\pi^2} +\mathcal{O}\left(\lambda^2\right)~.
\ee
Finally, inserting these results in eqs.~(\ref{4eq}) gives
\ba
0 &=&  \frac{\d}{\d m^2}\left[\bar{\Gamma}^{(2,0)}_{\theta,m}\right]_\lambda~,
\qquad\qquad\qquad
0= \frac{\d}{\d m^2}\left[\bar{\Gamma}^{(2,0)}_{\theta,M}\right]_\lambda ~,
\nn\\
0&=&  \frac{\d}{\d M^2}\left[\bar{\Gamma}^{(2,0)}_{\theta,m}\right]_\lambda~,
\qquad\qquad\qquad
0 =  \frac{\d}{\d M^2}\left[\bar{\Gamma}^{(2,0)}_{\theta,M}\right]_\lambda ~.
\ea
This is again a direct generalisation of the one-field result
\be
0 = \frac{\d}{\d m^2} \left[ \bar{\Gamma}^{(2)}_\theta\right]_\lambda ~. 
\ee
All in all, we have to conclude that the order $\lambda$ contribution to both $\bar{\Gamma}^{(2,0)}_{\theta,m}$ and $\bar{\Gamma}^{(0,2)}_{\theta,M}$ vanishes:
\ba
\bar{\Gamma}^{(2,0)}_{\theta,m} &=& 1+\mathcal{O}\left(\lambda^2\right)~,\nn\\
\bar{\Gamma}^{(2,0)}_{\theta,M} &=& \mathcal{O}\left(\lambda^2\right)~. \label{reeds}
\ea
Unfortunately, this discussion turns out very dry, just because in $\lambda \phi^4$ theory there is no external momentum appearing in the one-loop corrections to the two-point functions. But it is clear that the machinery works: we get equations for the derivatives with respect to $m^2 $ and $M^2$ of both $\bar{\Gamma}^{(2,0)}_{\theta,m}$ and $\bar{\Gamma}^{(2,0)}_{\theta,M}$, and in the meantime, we find four of the eight $\gamma_{\theta}$'s. Clearly, repeating the same procedure for $(n=0, N=2)$ will give equations for  both derivatives of $\bar{\Gamma}^{(0,2)}_{\theta,m}$ and $\bar{\Gamma}^{(0,2)}_{\theta,M}$ and produce the other four $\gamma_\theta$'s. It is all consistent.

\subsubsection{$\bar{\Gamma}^{(2,0)}$ }

We begin from the ``light" CS equation, eq.~(\ref{clight}), for $n=2,N=0$:
\be
2i\left[m^2\left(1+\gamma_{\phi,m}\right)\bar{\Gamma}^{(2,0)}_{\theta,m} + M^2 \gamma_{\Phi,m}\bar{\Gamma}^{(2,0)}_{\theta,M}\right] = \left[2m^2\frac{\d}{\d m^2}+\sum_i\beta_{\lambda_i,m}\frac{\d}{\d \lambda_i}+2\cdot \gamma_{\phi,m}\right]\bar{\Gamma}^{(2,0)} ~ . \label{clight2}
\ee
Inserting the results that we just found in eqs.~(\ref{reeds}), we see that at order $\lambda^0$ this equation reads
\be
2im^2 \cdot 1 = 2m^2\frac{\d}{\d m^2}\left(i\left[k^2+m^2\right]\right)~,
\ee
which is just another consistency test.

At order $\lambda$ we get
\be
2im^2 \left[\gamma_{\phi,m}\right]_\lambda = 2m^2\frac{\d}{\d m^2}\left[\bar{\Gamma}^{(2,0)}\right]_\lambda +2 \left[\gamma_{\phi,m}\right]_\lambda \cdot i\cdot (k^2+m^2)~. \label{20light}
\ee
Evaluating at $k^2=0$ only gives a consistency check (we have used this earlier already in order to derive the light CS equation):
\be
2im^2\left[\gamma_{\phi,m}\right]_\lambda =2 \left[\gamma_{\phi,m}\right]_\lambda \cdot i\cdot m^2~.
\ee

To get new information, we differentiate eq.~(\ref{20light}) with respect to $k^2$ and evaluate that at $k^2=0$. Then we can use the renormalisation condition $\left[\frac{d}{d k^2}\bar{\Gamma}^{(2,0)}\right]_{k^2=0} = i$. That yields
\be
0=2m^2\frac{\d}{\d m^2}\left[i\right] + 2\left[\gamma_{\phi,m}\right]_\lambda \cdot i \qquad \Rightarrow \qquad \left[\gamma_{\phi,m}\right]_\lambda=0~.\label{2mgam}
\ee
(In passing by we note that combining this result with eq.~(\ref{us2}) gives, as expected, $\left[\beta_{\lambda_\phi,m}\right]_{\lambda^2}=\frac{3 \lambda_\phi^2}{16\pi^2}$. In other words: the $(\theta,m)$ operation precisely yields the part of $\lambda_\phi$'s beta-function caused by the field of mass $m$ running in the loop.)

The result of eq.~(\ref{2mgam}) goes back into eq.~(\ref{20light}), and we get
\be
\frac{\d}{\d m^2}\left[\bar{\Gamma}^{(2,0)}\right]_\lambda=0~. \label{gamM2}
\ee

Now for Callan's ``heavy" equation, eq.~(\ref{cheavy}), still for $n=2$ and $N=0$:
\be
2i\left[M^2\left(1+\gamma_{\Phi,M}\right)\bar{\Gamma}^{(2,0)}_{\theta,M} + m^2 \gamma_{\phi,M}\bar{\Gamma}^{(2,0)}_{\theta,m}\right] = \left[2M^2\frac{\d}{\d M^2}+\sum_i\beta_{\lambda_i,M}\frac{\d}{\d \lambda_i}+2\cdot \gamma_{\phi,M}\right]\bar{\Gamma}^{(2,0)}~.
\ee
Again, there is no contribution to this equation at order $\lambda^0$.

At order $\lambda$ we get, using eqs.~(\ref{reeds})
\be
2im^2 \cdot \left[\gamma_{\phi,M}\right]_\lambda = 2M^2\frac{\d}{\d M^2}\left[\bar{\Gamma}^{(2,0)}\right]_\lambda+2\left[\gamma_{\phi,M}\right]_\lambda \cdot i\left(k^2+m^2\right)~. \label{orderl}
\ee
Evaluating this equation at $k^2=0$ and using the renormalisation condition $\bar{\Gamma}^{(2,0)}(0)=i$ up to all orders only gives a consistency check, again,
\be
2im^2 \cdot \left[\gamma_{\phi,M}\right]_\lambda   =  2\left[\gamma_{\phi,M}\right]_\lambda\cdot i m^2~,
\ee
Differentiating eq.~(\ref{orderl}) with respect to $k^2$ and evaluating at $k^2=0$, using  the renormalisation condition $\left[\frac{d}{d k^2}\bar{\Gamma}^{(2,0)}\right]_{k^2=0} = i$ does give something new, again:
\be
0=2M^2\frac{\d}{\d M^2}\left[i\right] + 2\left[\gamma_{\phi,M}\right]_\lambda \cdot i \qquad \Rightarrow \qquad \left[\gamma_{\phi,M}\right]_\lambda=0~.
\ee
With all that, eq.~(\ref{orderl}) becomes
\be
\frac{\d}{\d M^2}\left[\bar{\Gamma}^{(2,0)}\right]_\lambda=0~.
\ee
Together with eq.~(\ref{gamM2}), it is clear that we have
\be
\left[\bar{\Gamma}^{(2,0)}\right]_\lambda=0~. \label{res2}
\ee
Therefore, we have again not only found the above result, but in the process, we also got information on $\gamma_\phi$: $ \left[\gamma_{\phi,m}\right]_\lambda = \left[\gamma_{\phi,M}\right]_\lambda=0$. Clearly, the computation of $\bar{\Gamma}^{(0,2}$ will produce the order $\lambda$ contribution to the last remaining quantities  $\gamma_{\Phi,m}$ and $\gamma_{\Phi,M}$. We stress that the results in eqs.~(\ref{res1}) and (\ref{res2}) precisely coincide with the results obtained from the standard ``divergent" approach to compute loop corrections to correlation functions. 

Therefore, we conclude that the computation of the one-loop corrections to all ``divergent" correlation functions leaves one with all necessary information (i.e., all $\beta$'s, $\gamma$'s and $\gamma_\theta$'s) to move on to the two-loop corrections. The CS renormalisation scheme works recursively. We now need to compute all (manifestly convergent) two-loop $\Gamma^{(4)}$ and $\Gamma^{(2)}_{\theta\theta}$ diagrams. Meanwhile, by the usual skeleton expansion, we can address all one loop diagrams contributing to $\Gamma^{(4)}$ and $\Gamma^{(2)}_{\theta\theta}$ with subdivergences: we just insert one loop results for $\Gamma^{(2)}$ and $\Gamma^{(4)}$ obtained in the first step of the CS procedure. From the fact that the boundary conditions in eqs.~(\ref{rcon2}) hold up to all orders in the coupling constants, and that all correlation functions are analytic in $k^2$, we understand that this procedure can be repeated \emph{ad infinitum}. No matter what loop order we work, the following conclusion 
%from our companion paper \cite{MishaSander} 
always holds:  the pole mass of $\phi$ receives only very strongly suppressed (proportional to negative powers of $M$) contributions from the dynamics involving the heavy field $\Phi$. 

To iterate: we have used the CS method to arrive at the same results for the two- and four-point functions that one finds in the usual mass-dependent schemes. In both methods, the boundary/renormalisation conditions are taken at $k^2=0$. Therefore, by construction one finds that at external momenta $k^2 \ll M^2$, the influence of the heavy field on the light field's two-point function is heavily suppressed. (Again, this happens in both methods: they lead to the same results.) At higher external momentum scales $k^2 \propto M^2$ quantum corrections proportional to $M^2$ appear again. This, however, is not relevant for the experimentalist who has access only to external momentum scales $k^2\propto m^2 \ll M^2$. As before, all these results and their interpretation apply to both methods. However, in the traditional methods, large order $M^2$ cancellations between bare parameters and loop corrections are required to arrive at these results. We see, therefore, that in the CS method, the heavy particles completely decouple from the world of light particles and small momenta. This is valid for all the quantities, including the mass of the light particles.

\section{Conclusions}

We have provided a two-field generalisation of the Callan-Symanzik method of finite renormalisation (equivalent to the conventional zero external momentum subtraction scheme) of $\lambda \phi^4$ theory. We have established the two-by-two matrix equations in (\ref{bijnakl}) and (\ref{matr2}), and the corresponding boundary/renormalisation conditions of eqs.~(\ref{tree2f}), (\ref{rcon2}) and (\ref{bgtM2}). From these, all correlation functions of any combination of the two fields can be computed without running into intermediate UV divergences. We have provided an explicit computation of the two- and four-point correlation functions of one of the two scalar fields. The computation of all other correlation functions proceeds along precisely the same lines. A further generalisation to $n$ interacting scalar fields is trivial. 

As for an application of this work, the two-field model that we considered can act as a toy model for theories with widely separated mass scales. We see that in particular, the low energy sector of such a theory, including the light particle masses, is not affected by UV physics, confirming and extending the Appelquist-Carazzone theorem.

A suggestion for further work could be to establish equations that allow for a finite computation of zero-point functions in two-field models. In particular, we will need to compute $\bar{\Gamma}_{\theta\theta}$'s from manifestly  $\bar{\Gamma}_{\theta\theta\theta}$'s through a third CS equation. This seems to be technically challenging. We have seen that in the case of two fields, the first CS equation is an array of two equations, while the second one takes the form of a two-by-two matrix equation. We, therefore, expect the third CS equation to turn out as a ``two-by-two-by-two" matrix equation, containing eight independent equations. We have decided to postpone this analysis to later work.

\section*{Acknowledgements}

This work was supported by ERC-AdG-2015 grant 694896, by the Swiss National Science Foundation Excellence grant 200020B\underline{ }182864, and by the Generalitat Valenciana grant PROMETEO/2021/083.

\end{document}